\documentclass[
    reprint,
    preprintnumbers,
    superscriptaddress,
    nofootinbib,
     amsmath,amssymb,
     aps,
     prd,
    floatfix, longbibliography
    ]{revtex4-2}

    \usepackage{graphicx}
    \usepackage[caption=false]{subfig}
    \usepackage[usenames,dvipsnames]{xcolor} 
    \usepackage{pgfplots}
    \usepackage[utf8]{inputenc}
    \usepackage{color}
    \usepackage{hyperref}
    \usepackage[normalem]{ulem} 
    \usepackage{physics}
    \usepackage{enumitem}
    \usepackage{comment}
    \usepackage{bm}
\hypersetup{
  breaklinks = true,
  colorlinks   = true, 
  urlcolor     = Blue, 
  linkcolor    = Blue, 
  citecolor   = Blue 
}
    \allowdisplaybreaks[1]
    \graphicspath{{figures/}}

\usepackage{float}

\newcommand{\oxford}{Astrophysics, University of Oxford, DWB, Keble Road, Oxford OX1 3RH, United Kingdom}

\begin{document}

\title{Underdetermination of dark energy}

\author{William J. Wolf}
\email{william.wolf@stx.ox.ac.uk}
\affiliation{\oxford}
\author{Pedro G. Ferreira}
\email{pedro.ferreira@physics.ox.ac.uk}
\affiliation{\oxford}

\begin{abstract}
There is compelling evidence that the Universe is undergoing a late phase of accelerated expansion. One of the simplest explanations for this behaviour is the presence of dark energy.  A plethora of microphysical models for dark energy have been proposed. The hope is that, with the ever increasing precision of cosmological surveys, it will be possible to precisely pin down the model. We show that this is unlikely and that, at best, we will have a phenomenological description for the microphysics of dark energy. Furthermore, we argue that the current phenomenological prescriptions are ill-equipped for shedding light on the fundamental theory of dark energy. 
\end{abstract}

\maketitle


\section{Introduction}

Evidence for the accelerating expansion of the universe has been steadily accumulating since its discovery in 1998 \cite{SupernovaSearchTeam:1998fmf, SupernovaCosmologyProject:1998vns, BOSS:2016wmc, 2013ApJS..208...19H, Planck:2018vyg, DES:2018paw, Planck:2015bue}. While these observations are still consistent with a cosmological constant, there are other proposals that seek to account for this accelerating expansion by postulating the existence of dark energy \cite{Amendola:2015ksp}. Dark energy typically refers to an exotic form of matter, represented by additional fundamental field(s) that dynamically evolve over time and may (or may not) modify gravity on cosmological scales \cite{Clifton:2011jh}, depending on the exact nature of the field.

Characterizing this accelerating expansion usually begins by defining the equation of state.
\begin{equation}\label{DEeq}
w \equiv \frac{P}{\rho},
\end{equation}
where $P$ is the pressure and $\rho$ is the energy density of whatever is sourcing the accelerating expansion. If expansion is driven by a cosmological constant, the equation of state is locked in at the value $w=-1$. If, on the other hand, expansion is driven by a dynamical field, it will evolve over time and the equation of state can then be described as a function of $a$, the scale factor of the universe, so that $w=w(a)$.

\begin{figure}
    \centering
    \includegraphics[width=0.4\textwidth]{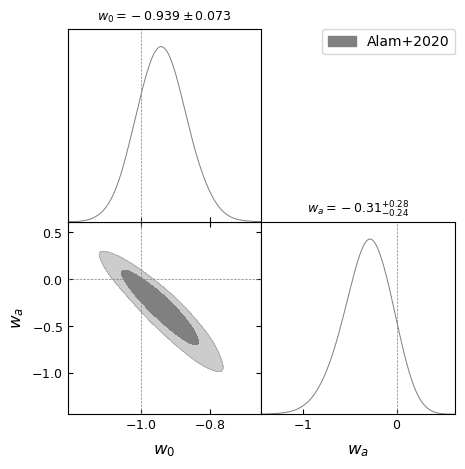}
    \caption{Current constraints on the $(w_0, w_a)$ parameter space. Broadly speaking, this parameter space can be divided up into a few regions. ``Freezing" quintessence corresponds to the region of the parameter space where $w_a > 0$ and refers to models where the dark energy equation of state is evolving asymptotically towards $w_{DE} \simeq -1$, while ``thawing" quintessence corresponds to the region where $w_a < 0$ and refers to models where the dark energy equation of state is evolving away from $w_{DE} \simeq -1$ to less negative values. The region given by $w<-1$ is known as the phantom region. It requires more exotic physics to describe and will not be a focus of this study. As we can see, most of the viable parameter space is locating in the thawing region.}
    \label{paramspace}
\end{figure}

Ultimately, one would like to be able to extract the detailed time dependence of the equation of state in the form of $w(a)$ or $w(z)$ where $z$ is the redshift, $1+z\equiv a^{-1}$ and, indeed, there have been attempts at doing so \cite{Said:2013jxa,Raveri:2021dbu}. In practice there is limited information that one can extract about the evolution of $w(a)$ and, as a result, this has led to the wide-spread adoption of a very natural and popular parameterization known as the  Chevallier-Polarski-Linder (CPL) parameterization \cite{Linder:2002et, Chevallier:2000qy}, given by:
\begin{equation}\label{param}
w(a)=w_0+w_a(1-a).
\end{equation}
This parameterization approximates the dark energy equation of state close to today ($a\sim1$); where $w_0$ gives the value of the equation of state now and 
$w_a$ characterizes the temporal evolution of dark energy. This allows for a very useful description of dark energy models in terms of the  parameters $(w_0,w_a)$. Fig.~(\ref{paramspace}) shows the current constraints on these parameters. A cosmological constant, $\Lambda$, is of course given by $w_a = 0$ and $w_0 = -1$, while various dynamically driven dark energy possibilities occupy the rest of the parameter space. 

One interesting thing to note is that, while a cosmological constant is still favoured, there is a tremendous amount of open parameter space and future work that must be done in order to measure $(w_0,w_a)$. This invites the following question: what can pinning down the values of $(w_0,w_a)$ teach us about the microphysical nature of dark energy? As others have noted \cite{Caldwell:2005tm}, constraining down to $w_a \sim 0$ and $w_0 \sim -1$ strongly points towards a cosmological constant, but can never fully eliminate dynamically driven dark energy as a possibility because it may simply not yet have entered a stage of temporal evolution that is observable to us. There is of course another possibility. What if future surveys indicate that $w_0 \neq -1$ and that there is significant temporal evolution captured in $w_a$? 

Clearly, observing $w_a \neq 0$ with statistical significance would rule out the basic cosmological constant scenario and point towards the existence of dark energy driven expansion. However, would such observations allow us to definitively say anything more specific about dark energy, other than inferring its dynamical nature? Most obviously, we would like to gain an understanding of the fundamental microphysics responsible of dark energy \cite{Marsh:2014xoa,Garcia-Garcia:2019cvr, Traykova:2021hbr, Linder:2023klx, Raveri:2017qvt, Peirone:2017lgi, Colgain:2021pmf, Caldwell:2005tm, Park:2021jmi, Escamilla:2023oce, Huterer:2006mv, Scherrer:2015tra}. This presumably (in keeping with the field theoretic paradigm of modern physics) would be captured in a Lagrangian expression; one that includes structural information concerning the relevant types of dynamical field(s), the couplings to gravity and other material fields, and the forms of the respective kinetic and potential terms. To this end, the shear volume of dark energy models that have been explored is extensive: a by no means exhaustive list includes canonical quintessence, k-essence, $\alpha$-attractors, f(R) gravity, Horndeski scalar-tensor gravity, DHOST theories, Einstein-Aether theories, bi-metric theories, and many more \cite{Clifton:2011jh, Amendola:2015ksp, Joyce:2016vqv, Tsujikawa:2013fta, Garcia-Garcia:2018hlc, Sotiriou:2008rp, Kobayashi:2019hrl, Kase:2018aps, Copeland:2006wr, Charmousis:2011bf, Horndeski:1974wa}. Furthermore, many of these theories have a large number of specific realizations. 

It is hard to overstate the value of gaining such access to the fundamental microphysics of dark energy. For example, if dark energy is caused by some exotic field, this could allow us to situate dark energy within the standard model of particle physics. Such information might give us crucial information regarding persistent conundrums such the Hierarchy problem or what further symmetry principles are at play in particle physics \cite{Carroll:1998zi, Martin:2008qp}. If modifications to gravity are at play, this could give us information regarding gravity's renormalizability \cite{Sotiriou:2008rp}. Cosmologically speaking, it could indicate the ultimate fate of the universe itself: will the universe continue in an accelerating expansion forever (as in the cosmological constant scenario or particular ``freezing" realizations of dark energy) or will the universe stop accelerating (as in ``thawing" realizations of dark energy) \cite{Caldwell:2005tm}; or could it even potentially begin contracting \cite{Andrei:2022rhi, Steinhardt:2001st}? 

In this article, we will attempt to shed some further light concerning the degree to which constraining and determining the values of the parameters $w_0$ and $w_a$ will inform us about the fundamental microphysics driving dark energy. Our verdict is largely pessimistic: it is unlikely that constraining the $(w_0, w_a)$ plane will ever allow us to single out a specific theory of dark energy, even if we were to detect clear evidence of dynamical behavior ($w_0 \neq -1$ and $w_a \neq 0$). 

We support our argument in two primary ways: (i) we demonstrate that there is significant underdetermination in the $(w_0, w_a)$ plane in the sense that there is a complete degeneracy between multiple realizations of dark energy and their mapping over vast swaths of this parameter space; (ii) we highlight some shortcomings of using $(w_0, w_a)$  by demonstrating that this parameterization is remarkably sensitive to the properties of the data used to constrain it, introducing additional confounding factors.

On (i), it has been previously suggested in many places in the literature that typical realizations of ``freezing" and ``thawing" models of dark energy occupy small, well-defined regions in the $(w_0, w_a)$ plane; and that measurements of these parameters would clearly indicate whether or not dark energy was described by a simple realization (e.g.\ single field, minimally coupled, canonical, etc.) 
of one of these classes \cite{Linder:2007wa, Barger:2005sb, Caldwell:2005tm, Clemson:2008ua, Linder:2006sv}. Furthermore, the typical understanding holds that finding values for $w_0$ and $w_a$ outside these narrow regions would indicate more complicated dynamics, exotic physics, or highly unnatural fine-tuning. 

We challenge this orthodoxy specifically regarding thawing dark energy. We do so by exploring arguably the simplest model of dark energy, a minimally coupled scalar field with a quadratic potential. We analytically demonstrate, using exact solutions in a matter-dominated background, that this model can arbitrarily sweep across huge sections of the $(w_0, w_a)$ plane depending on simple choices of model parameters. Furthermore, this model is of particular interest because it can be understood as an effective field theory approximation of a significant number of other distinct models \cite{Burgess:2020tbq}; this means that it provides a map between any space on the $(w_0, w_a)$ plane that it sweeps and many distinct models. We then show that these conclusions hold more generally where we numerically integrate the equations for a mixed dark matter/dark energy universe thought to describe the actual universe we live in. Indeed, the difficulty of constructing a unique potential from observational data has been noted in other places (see e.g.\ \cite{Park:2021jmi}). Here we analytically and numerically illustrate that determining a unique dark energy is almost certainly impossible by using this simple model (and the many models that it approximates) to cover significant portions of the remaining viable parameter space.

Regarding (ii), we demonstrate that the $w_0, w_a$ parameterization is very sensitive to the range of redshifts one fits over. That is, we show that this parameterization captures some $w(a)$ evolutions for the quadratic model reasonably well, but fairs far less successfully with others; making the mapping between theoretical dark energy models and the $(w_0, w_a)$ phase space sensitively dependent on arbitrary choices regarding the range of redshifts one fits over. We illustrate this effect with a few different choices of survey parameters. 

The paper proceeds as follows. Sect.~(\ref{sec:quintessence}) provides an overview of quintessence: we introduce two flavours of the quadratic potential (``slow-roll" and ``hilltop" variations), derive their analytic solutions, and discuss how this model can effectively approximate a huge variety of models with distinct potentials. Sect.~(\ref{sec:DEplane}) derives analytic expressions relating the quadratic potential to the $(w_0, w_a)$ parameter space and demonstrates that this model can sweep huge portions of the $(w_0, w_a)$ plane. We then numerically solve the equations of motion and show that this conclusion holds in a realistic universe with both dark matter and dark energy. Sect.~(\ref{sec:surveys}) discusses these results in light of current surveys aimed at constraining $w_0$ and $w_a$. Sect.~(\ref{sec:conclusion}) concludes.

\section{Quintessence}\label{sec:quintessence}
Quintessence is perhaps the simplest theoretical proposal for dark energy (see e.g.\ \cite{Peebles:1987ek,Caldwell:1997ii} for early papers on the subject), and is characterized by a dynamical scalar field $\varphi$ with a canonical kinetic term and a potential $V(\varphi)$. The theory is given by an action of the following form:
\begin{equation}
S=\int d^4 x \sqrt{-g}\left[\frac{1}{2} M_{\mathrm{pl}}^2 R-\frac{1}{2} g^{\mu \nu} \partial_\mu \varphi \partial_\nu \varphi-V(\varphi)\right]+S_m,
\end{equation}
where $M_{pl}$ is the reduced Planck mass, $g$ is the determinant of the metric $g_{\mu\nu}$, $R$ is the Ricci curvature scalar, and $S_m$ is the action for matter. The scalar field minimally couples to gravity through the metric determinant and is assumed to have no direct coupling to matter fields. 

The equation of state for dark energy $w_{DE}$ driven by such a field is given by:
\begin{equation}\label{DEeq}
w_{DE} \equiv \frac{P_\varphi}{\rho_\varphi}=\frac{\dot{\varphi}^2 / 2-V(\varphi)}{\dot{\varphi}^2 / 2+V(\varphi)},
\end{equation}
where $P_{\varphi}$ is the pressure and $\rho_{\varphi}$ is the energy density of the scalar field $\varphi$. Here, we can see that $w_{DE}$ will dynamically evolve in time with the evolution of the scalar field. 

The evolution of the scalar field is given by the scalar field equation of motion:
\begin{equation}\label{ScalarEOM}
    \ddot{\varphi} + 3 H \dot{\varphi} + V'(\varphi) = 0,
\end{equation}
where $V'(\varphi) = dV/d\varphi$ and $H$ gives the expansion rate of the universe through the first Friedmann equation:
\begin{equation}\label{Friedmann}
H^2=\left(\frac{\dot{a}}{a}\right)^2=\frac{1}{3M^2_{pl}} (\rho+\rho_{\varphi}),
\end{equation}
where $a$ is the scale factor of the universe and $\rho$ is the density of matter and radiation. Together, these equations completely determine the dynamics of quintessence. Solving for the dynamics of $\varphi$ allows one to then determine the evolution of $w_{DE}$ through Eq.~(\ref{DEeq}).

The most extensively studied quintessence models fall into two broad categories: \textit{freezing} quintessence or \textit{thawing} quintessence. As the names suggest, freezing quintessence describes dark energy evolution $w_{DE}$ which was different from $w_{DE} \simeq -1$ in the past, but is now evolving asymptotically towards this value as the universe expands (i.e.\ ``freezing in") and broadly speaking falls in the $w_a > 0$ part of the parameter space; thawing quintessence describes dark energy evolution where $w_{DE}$ has been close to $w_{DE} \simeq -1$ in the past, but is now beginning to evolve towards larger values as the universe expands (i.e.\ ``thawing out") and broadly speaking falls in the $w_a < 0$ region of the parameter space. Here, we will be particularly concerned with thawing models as this is the less constrained part of the parameter space. 

So far everything has been completely general. One must then specify a form of the potential function $V(\varphi)$, which will enable one to solve the equations of motion, as well as give a particular realization of the relevant microphysics driving dark energy and allow for a direct investigation of any observables associated with the specific model(s). Mirroring a similar situation with inflation \cite{Martin:2013tda, Sousa:2023unz}, the number of distinct models that are distinguished by their potential function is  large. The possible forms of the potential include all imaginable varieties and combinations of power laws, inverse power laws, exponentials, axions, trigonometric functions, and many more. See e.g.\ \cite{Martin:2008qp, Tsujikawa:2013fta, Amendola:2015ksp} for reviews on quintessence which mention many of these potentials and \cite{Martin:2013tda} for an exhaustive review on inflationary potentials, many of which can be similarly adapted to quintessence. Furthermore, the exact form of the potential holds important information regarding the microphysics responsible for dark energy, the overarching structure and integration of such a field into the standard model of particle physics, and the ultimate fate of the universe's evolutionary trajectory.

However, in this article, we will only consider arguably the simplest model of quintessence: dark energy driven by a quadratic $m^2 \varphi^2$ potential. The reason for this is that this model can be understood from an effective field theory perspective as the leading order expansion of many other distinct models. In other words, an arbitrary (analytic) single-field model can be represented by an expansion:
\begin{equation}\label{expand}
V=V_0+\left.\frac{d V}{d \varphi}\right|_{\varphi=0} \varphi+\left.\frac{1}{2} \frac{d^2 V}{d \varphi^2}\right|_{\varphi=0} \varphi^2+\left.\frac{1}{6} \frac{d^3 V}{d \varphi^3}\right|_{\varphi=0} \varphi^3+\cdots.
\end{equation}
However, this can often be cast in a form that resembles the quadratic potential. For example, consider another well-studied model described by an exponential potential $V(\varphi) = V_0e^{-\lambda \varphi}$ \cite{Caldwell:1997ii, Rubano:2001su, Barreiro:1999zs, Copeland:1997et}. Expanding, we have that,
\begin{equation}
    V(\varphi) = V_0\left(1 - \lambda \varphi + \frac{\lambda^2}{2}\varphi^2 ...\right).
\end{equation}
However, a simple field re-definition $\varphi \rightarrow \varphi - \varphi_0$ where $\varphi_0$ is constant allows us to cast this in a quadratic form resembling the $m^2$ potential. That is, 
\begin{align}
    V(\varphi - \varphi_0) = V_0 \left(1 + \frac{\lambda^2}{2} (\varphi^2) - \varphi (\lambda + \lambda^2 \varphi_0) + ...\right),
\end{align}
where $\varphi_0$ can be chosen such that $\varphi_0 = -1/\lambda$ and the linear term vanishes, and any constant terms can be absorbed into the definition of $V_0$. Thus, we see that an exponential potential can, under certain circumstances, be well-approximated as a quadratic potential of the form $V(\varphi) \simeq V_0 + \frac{1}{2}m^2\varphi^2$, where in this case the ``effective mass" is given by $ \lambda^2 V_0 = V''$. Similarly, one could consider the potential of pseudo Nambu-Goldstone boson given by $V(\varphi)=V_0[\cos (\varphi / f)+1]$ \cite{Dutta:2008qn, Abrahamse:2007te, Frieman:1995pm} or the supergravity motivated potential $V(\varphi)=V_0(2-\cosh \sqrt{2} \varphi)$ \cite{Kallosh:2002gg}, and find that under this expansion these models are also approximated by a quadratic potential with a negative sign in front of the mass term $V(\varphi) \simeq V_0 - \frac{1}{2}m^2\varphi^2$. Given that this is an effective parameterization in a local region of the potential, we do not need to concern ourselves with the fact that $V(\varphi)$ may be unbounded from below.

If the $m^2 \varphi^2$ model can cover large swaths of the parameter space in dark energy observables, this also implies a significant underdetermination in the form of the potential and the underlying microphysics driving dark energy. That is, any dark energy trajectories that can be obtained by an $m^2 \varphi^2$ model can also trivially be recast as any number of other dark energy models (admitting such an expansion of course) with distinct functional forms of their potentials and different fundamental microphysics. As indicated above, in the following we will consider two variations of the quadratic model: (i) the positive $m^2 \varphi^2$ model, which is representative of the standard slow-roll thawing quintessence models and (ii) the negative $m^2 \varphi^2$ model, which is representative of so-called ``hilltop" quintessence models. Let us now flesh out our understanding of these two branches of the quadratic potential.

\subsection{$V_0+\frac{1}{2}m^2 \varphi^2$}

We begin with the positive quadratic model as this is representative of the most standard kind of thawing quintessence. Standard thawing models are characterized by a set of ``slow roll" conditions (not to be confused with the inflationary slow roll conditions) \cite{Scherrer:2007pu}. They are:
\begin{equation}\label{slowroll1}
\left(\frac{1}{V} \frac{d V}{d \varphi}\right)^2 \ll 1,
\end{equation}
\begin{equation}\label{slowwoll2}
\frac{1}{V} \frac{d^2 V}{d \varphi^2} \ll 1.
\end{equation}
These conditions are important because they ensure that the potential is flat enough to yield $w_{DE} \simeq -1$ at early times, before the field starts slowly rolling, driving $w_{DE}$ to larger values at late time. 

As mentioned earlier, our representative of standard, slowly rolling thawing models will be given by the quadratic potential:
\begin{equation}\label{+mV0}
    V(\varphi) = V_0 + \frac{1}{2}m^2\varphi^2.
\end{equation}
Another benefit of this particular model is that, depending on the dominant background energy component, Eqs.~(\ref{ScalarEOM}-\ref{Friedmann}) can be solved exactly (see e.g.\ \cite{Liddle:1998xm, Scherrer:2022umm, Marsh:2010wq}). Here and throughout this article, we work in dimensionless variables $t \rightarrow H_0 t$, $H \rightarrow H / H_0, \varphi \rightarrow \varphi / m_{p l}, m \rightarrow m / H_0$, where $H_0$ is the Hubble constant today. We also take the ansatz $a(t) \propto t^p$ which is true in both radiation and matter domination, leading to exact analytic expressions for $\varphi$: 
\begin{equation}
\varphi(t)=a(t)^{-3 / 2}(m t)^{1 / 2}\left(A J_n(m t)+B Y_n(m t)\right),
\end{equation}
where $J_n$ and $Y_n$ are Bessel functions of the first and second kinds respectively and $n=(1 / 2) \sqrt{9 p^2-6 p+1}$. To simplify, we work in a matter dominated background ($p=2/3, n=1/2$). As we will show later though, our results are generalisable to the accelerated expansion era. The general solution for $\varphi (t)$ is of the form:
\begin{equation}
    \varphi(t)=\frac{A}{t} \sin (m t)+\frac{B}{t} \cos (m t).
\end{equation}
The particular solution $\varphi(t)$ for the initial value problem with an initial value $\varphi_i$, an initial velocity $\dot{\varphi}_i$, at some early initial time $t_i$ (where we have written this using the small parameter $\xi = m t_i \ll 1$) is:
\begin{equation}
\varphi(t)=\frac{\varphi_i}{mt}\left[\sin (mt-\xi)+\xi \cos (mt-\xi)\right]+\frac{\dot{\varphi_i} \xi}{m^2 t} \sin (mt-\xi),
\end{equation}
Expanding in $\xi$, to leading order we find $\varphi(t)$:
\begin{align}\label{phi+}
\begin{split}
    \varphi(t) &\simeq \frac{\varphi_i}{mt} \sin (mt), \\
   \varphi(t) &\simeq \varphi_i \left(1-\frac{(mt)^2}{6}\right),
\end{split}
\end{align}
where in the last line we have expanded again (this time in $mt$) as we have used that $mt < 1$ so that the scalar field has not entered the oscillatory regime (i.e.\ this region of the potential will not produce $w_{DE} \sim -1$). Eq.~(\ref{phi+}) will allow us to write down an analytic expression for Eq.~(\ref{DEeq}) for this quintessence model.

\subsection{$V_0-\frac{1}{2}m^2 \varphi^2$}

While the slow roll conditions given in Eqs.~(\ref{slowroll1}-\ref{slowwoll2}) are sufficient to generate a typical thawing quintessence model, they are not both strictly necessary. As described first in \cite{Dutta:2008qn}, one can relax the condition given in Eq.~(\ref{slowwoll2}) and still maintain the qualitative features of thawing quintessence, but with some notable differences. In this scenario, the scalar field rolls down a local maximum of the potential. Here, the model is such that the field remains close enough to the local maximum that Eq.~(\ref{slowroll1}) is still valid, but not necessarily Eq.~(\ref{slowwoll2}). This has been dubbed \textit{hilltop} quintessence \cite{Dutta:2008qn, Chiba:2009sj} in analogy with similar hilltop models of inflation (see e.g.\ \cite{Boubekeur:2005zm, Stein:2022cpk}).

We will examine a hilltop quintessence model given by a quadratic potential both because (as before) there are simple analytic expressions for the scalar field dynamics and this model approximates a tremendous variety of distinct dark energy models as their leading order expansion. The potential is given by:
\begin{equation}\label{-mV0}
    V (\varphi) = V_0 - \frac{1}{2}m^2 \varphi^2.
\end{equation}
Eqs.~(\ref{ScalarEOM}-\ref{Friedmann}) can again be solved exactly for matter and radiation dominated backgrounds \cite{Alam:2003rw}. The only difference is the presence of the minus sign in front of $m^2$; resulting in similar analytic solutions this time given by
\begin{equation}
\varphi(t)=a(t)^{-3 / 2}(m t)^{1 / 2}\left(A I_n(m t)+B K_n(m t)\right),
\end{equation}
where $I_n$ and $K_n$ are modified Bessel functions of the first and second kinds respectively. In the matter dominated era, the general solution for $\varphi$ is:
\begin{equation}
    \varphi(t)=\frac{A}{t} \sinh (m t)+\frac{B}{t} \cosh (m t).
\end{equation}
The particular solution $\varphi(t)$ for the initial value problem with an initial value $\varphi_i$, an initial velocity $\dot{\varphi}_i$, at some early initial time $t_i$ (where we have written this using the small parameter $\xi = m t_i \ll 1$) is:
\begin{equation}
\begin{split}
\varphi(t)&=\frac{\varphi_i}{mt}\left[\sinh (mt-\xi)+\xi \cosh (mt-\xi)\right]\\
&+\frac{\dot{\varphi_i} \xi}{m^2 t} \sinh (mt-\xi),
\end{split}
\end{equation}
Expanding in $\xi$, to leading order we find:
\begin{align}\label{phi-}
\begin{split}
    \varphi(t) &\simeq \frac{\varphi_i}{mt} \sinh (mt)
\end{split}
\end{align}
Notice that these are hyperbolic functions so we do not need to demand that $mt < 1$ to avoid the oscillatory regime. Similarly, Eqs.~(\ref{phi-}) will allow us to write down an analytic expression for Eq.~(\ref{DEeq}) for this hilltop model. Having obtained solutions for the scalar field, we will now proceed to analyzing the evolution of the dark energy equation of state in these respective models.


\section{Dark energy in the $(w_0, w_a)$ plane}\label{sec:DEplane}

\subsection{The $w_0$-$w_a$ parameterization revisited}

As mentioned earlier, the favoured parameterization for dark energy is given by:
\begin{equation}
w(a)=w_0+w_a(1-a). \nonumber
\end{equation}
Before we proceed, we need to be clear about how $w_0$ and $w_a$ are actually determined. In practice, a given choice of $w_0$ and $w_a$ will determine the time evolution of the dark energy density, $\rho_{\rm DE}$ or, in the case of quintessence, $\rho_\varphi$. This density, via the Friedmann equations, will determine the expansion rate of the Universe as a function of time. Given a set of observations -- typically measurements of standards candles over a redshift range -- one can pin down values of the expansion rate at different times, or redshift. One then finds the $w_0$ and $w_a$ (and their associated uncertainties) that best fit the observations. In other words, in practice, $w_0$ and $w_a$ arise from fitting the data over a range of redshifts. 

We can consider another parametrization of the equations of state,
\begin{eqnarray}
    w(a)={\tilde w}_0+{\tilde w}_a(1-a) \nonumber
\end{eqnarray}
where
\begin{eqnarray}
    {\tilde w}_0&\equiv&w(a=1) \nonumber\\
    {\tilde w}_a&\equiv&-\frac{dw}{da}(a=1) \nonumber
\end{eqnarray}
One can then calculate the values of $({\tilde w}_0, {\tilde w}_a)$ for different choices of potentials, background expansions and initial conditions. This can give us a useful, often analytic, understanding of their features and how they relate to the underlying theory. But it should be clear, from the outset that the values obtained in this way, while indicative, will not be the values one obtains through the fitting procedure described above. We will bear this in mind as we proceed in what follows and we will be particularly careful, by using this notation, to distinguish between the two different ``types" of ($w_0,w_a$).

Let us now explore how the dynamics for different dark energy models (as well as their mapping into the $({\tilde w}_0, {\tilde w}_a)$ parameters) can be straightforwardly understood from the scalar field equation of motion,
\begin{equation*}
 \ddot{\varphi} + 3 H \dot{\varphi}+V_{, \varphi}=0,   
\end{equation*}
where we have an ``acceleration" term,  a ``friction" term, and a ``potential" term.  

These terms will be important at different epochs depending on the classification of dark energy model. For example, in freezing models at early times the evolution of the potential term is significant as $w_{DE} \neq -1$. Yet, as a freezing model approaches $w_{DE} \simeq -1$, the potential becomes very flat and the dynamics are dominated by the friction term. The situation is different with thawing models as at early times the equation of state is locked in at $w_{DE} \simeq -1$ and the friction term is dominant. As a thawing model evolves away from $w_{DE} \simeq -1$, the potential term becomes more important as it is the field's evolution through its potential that causes dark energy to thaw. 

This leads to well-defined bounds that typical (i.e.\, slow-roll) freezing and thawing models respectively are thought to live in throughout the course of their evolution, where these bounds only span a tiny subset of the broader freezing ($w_a > 0$) or thawing ($w_a < 0$) regions. In particular, the thawing bounds are given by $-1 \lesssim  {\tilde w}_a/(1+{\tilde w}_0) \lesssim -3$  and it is commonly accepted in the literature that thawing quintessence models are constrained to live within this range (see e.g.\ \cite{Linder:2007wa, Barger:2005sb, Caldwell:2005tm, Clemson:2008ua, Linder:2006sv}); which clearly represents only a tiny fraction of the broader thawing region depicted in Fig.~(\ref{paramspace}). These bounds are largely determined by the background in which the scalar field evolves in. Essentially, one can examine the ratios between the friction and acceleration terms and the potential and acceleration terms, and conclude that a slow-roll thawing model will be one for which this lower bound obtains when in a radiation dominated background and this upper bound obtains when in a matter dominated background (see e.g.\ \cite{Linder:2006sv, Linder:2007wa, Caldwell:2005tm} for further discussion). Thus, a thawing quintessence model in a matter-dominated universe will consequently evolve along a narrow line given by ${\tilde w}_a/(1+{\tilde w}_0) \sim -3$. 

However, most of the available parameter space in the $({\tilde w}_0,{\tilde w}_a)$ plane clearly lies outside of this given range. The question is, can we find a way to use the $V_0\pm m^2 \varphi^2$ models to fill in any of the remaining parameter space outside of the typically quoted thawing bounds? The answer, as we shall see below, is yes. This is because hilltop models of the type considered above have notably different evolutionary trajectories than the more standard slow-roll thawing models \cite{Dutta:2008qn}. 

\subsection{Analytic expressions for $w(a)$}

\subsubsection{$V_0+\frac{1}{2}m^2\varphi^2$}

In order to determine the evolution of $w(a)$ for the positive quadratic model, we utilize Eqs.~(\ref{+mV0}, \ref{phi+}) and replace them in Eq.~(\ref{DEeq}) to get, 
\begin{align}
\begin{split}
w(t) \simeq -1 + \frac{m^2 \varphi_i^2}{9V_0} (mt)^2.
\end{split}
\end{align}
In order to determine the expression as a function of the scale factor, we recall that during matter domination $t=t_0 a^{3/2}$ with respect to some reference time $t_0$. 

To determine $({\tilde w}_0, {\tilde w}_a)$ we need, 
\begin{align}
    w(a) &= -1 + X (mt_0)^2 a^3, \\
    \frac{dw}{da} (a) &= 3 X (mt_0)^2 a^2,
\end{align}
where $X = \frac{m^2 \varphi_i^2}{9V_0}$. At $a=1$ and using Eq.~(\ref{param}), we find that,
\begin{equation}
    \frac{{\tilde w}_a}{1+{\tilde w}_0} = -3.
\end{equation}
As expected, the slow roll $m^2\varphi^2$ model evolves along this previously quoted line in the $({\tilde w}_0, {\tilde w}_a)$ plane with a slope of $-3$.

\subsubsection{$V_0-\frac{1}{2}m^2\varphi^2$}

Similarly, we use Eqs.~(\ref{-mV0}, \ref{phi-}, \ref{DEeq}) to determine $w(a)$ for the hilltop model.
\begin{align}
\begin{split}
w(t) & \simeq -1+\frac{(m \varphi_i)^2}{V_0 (mt)^4}\left(\sinh (m t)-m t \cosh (m t)\right)^2
\end{split}
\end{align}
Also utilizing $t=t_0 a^{3/2}$ and defining $\epsilon = \frac{(m \varphi_i)^2}{2V_0}$, we now have that\footnote{See \cite{Dutta:2008qn, Chiba:2009sj} for derivations of related, but more complicated expressions for $w(a)$ in hilltop models. The expressions derived there represent approximate analytic solutions for $w(a)$ assuming that the universe evolves in a dark energy dominated background. In this paper, we will focus on the exact analytic solution in a matter dominated background and the full numerical solution for a mixed dark matter/dark energy universe.}, 
\begin{align}
\begin{split}
    w(a) = -1 + \frac{2\epsilon}{a^{6}(mt_0)^4}\left[\sinh (m t_0 a^{3/2})\right. \\
    \left.-m t_0 a^{3/2} \cosh (m t_0 a^{3/2})\right]^2
\end{split}
\end{align}
and that, 
\begin{align}
\begin{split}
    &\frac{dw}{da}(a) = \frac{-6 \epsilon}{a^4 (mt_0)^4} \sinh \left( m t_0a^{3/2}\right) \left[\sinh \left( m t_0 a^{3/2}\right)\right. \\
    & \left.- m t_0a^{3/2} \cosh \left( m t_0 a^{3/2}\right)\right] \\
    &+\frac{-12\epsilon}{a^{7}(mt_0)^4}\left[\sinh (m t_0 a^{3/2})-m t_0 a^{3/2} \cosh (m t_0 a^{3/2})\right]^2 \\
\end{split}
\end{align}
At $a=1$ and using Eq.~(\ref{param}), the expression simplifies considerably, giving: 
\begin{equation}\label{a1hilltopslope}
\frac{{\tilde w}_a}{{\tilde w}_0+1}=6+\frac{3\left(m t_0\right)^2 \sinh \left(m t_0\right)}{\sinh \left(m t_0\right)-\left(m t_0\right) \cosh \left(m t_0\right)}.
\end{equation}
The behavior of the hilltop model will be notably different than that of the slow roll model. For example, the behavior for small $m$ (or equivalently $V'' \ll 1$) approaches ${\tilde w}_a \approx -3 (1+{\tilde w}_0)$. This is not surprising as this is the regime in which the hilltop model approximates the slow roll conditions in the previous model. However, as one violates these conditions with larger $m$, the trajectory of $w(a)$ can change substantially. 

A consequence of this is that these models can now look significantly different in the $({\tilde w}_0, {\tilde w}_a)$ plane and one can quite easily tune the slope to be steeper depending on the choice of $m$. For example, Eq.~(\ref{a1hilltopslope}) indicates that the slope  would span between $-3 \lesssim {\tilde w}_a/({\tilde w}_0+1) \lesssim -15.5$ for $.01 \leq mt_0 \leq 6$. This (i) indicates that we can use the quadratic hilltop model to essentially find any value we want in the $({\tilde w}_0, {\tilde w}_a)$ plane simply by selecting appropriate mass/$V''$ and $V_0$ parameters. Furthermore, due to this model's ability to approximate any other model that admits a Taylor expansion of the type considered in Eq.~(\ref{expand}), this (ii) similarly indicates that we can map any distinct dark energy model within this family (quadratic, exponentials, axions, etc) to any point in the $({\tilde w}_0, {\tilde w}_a)$ plane that the quadratic model can reach through any one of these models' ``effective" mass and $V_0$ terms. Taken together, these point to a potentially serious underdetermination in the microphysics of dark energy with respect to our observable parameterization of $w(a)$. It seems like no matter where we end up in the $({\tilde w}_0, {\tilde w}_a)$ plane within the broader thawing region, there are always a multitude of models that can be easily mapped to any point with a simple choice of parameters. 

\subsection{Covering the $({\tilde w}_0, {\tilde w}_a)$ plane}

\begin{figure}
    \subfloat[Slow-roll thawing quintessence trajectories.\label{slowrolltrajectory}]{\includegraphics[width=\columnwidth]{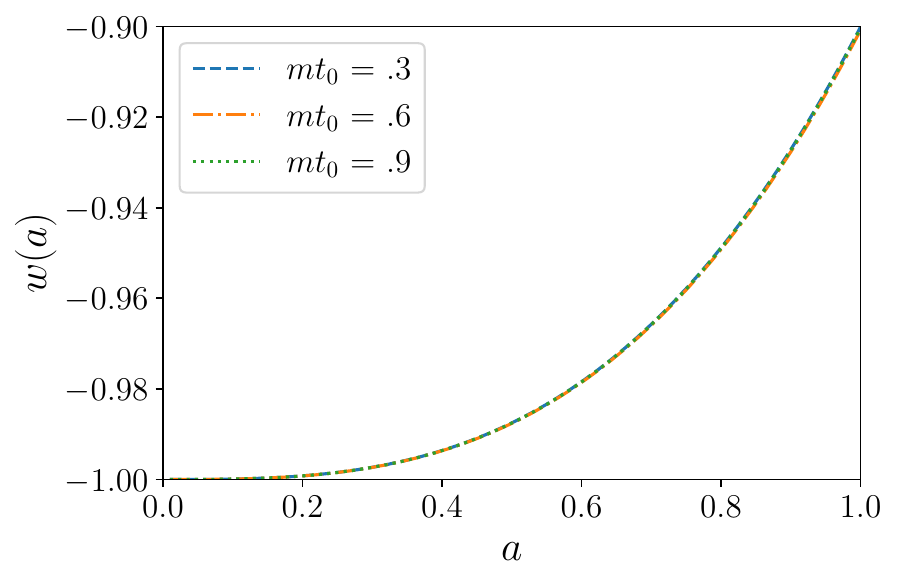}}
    \hfill
    \subfloat[Hilltop thawing quintessence trajectories.\label{hilltoptrajectory}]{\includegraphics[width=\columnwidth]{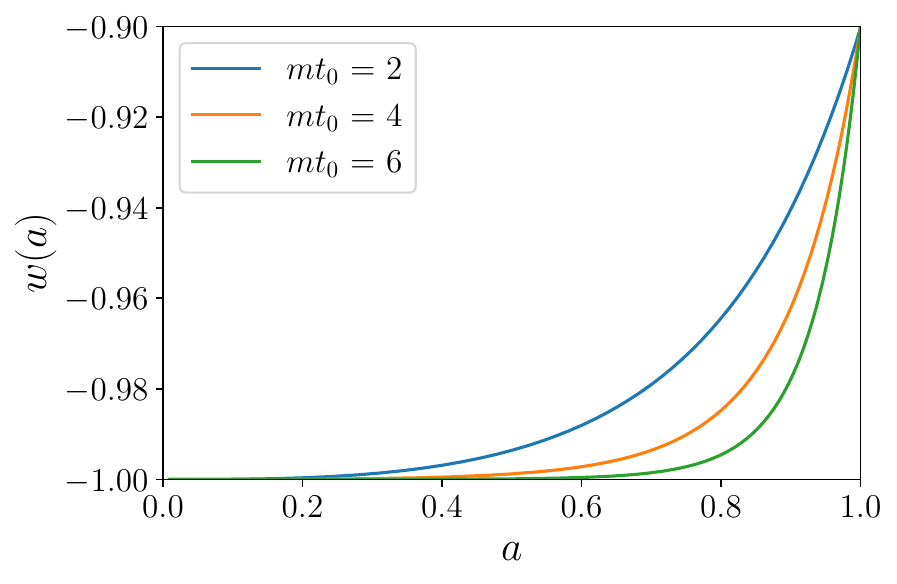}}
    \caption{$w(a)$ evolution for different choices of model parameters for the slow-roll/positive quadratic model and the hilltop/negative quadratic model. The parameter $mt_0$ was varied and $X$ was chosen so that ${\tilde w}_0 \simeq .90$. For the slow-roll models, all converge on the same evolutionary track and will have the same slope in the $({\tilde w}_0, {\tilde w}_a)$ plane. All the hilltop models have very different evolutions and the trajectories become steeper as $mt_0$ is increased. Consequently, they will be represented differently in the $({\tilde w}_0, {\tilde w}_a)$ plane.}
\end{figure}

Let us now see more concretely how these models map into the $({\tilde w}_0, {\tilde w}_a)$ plane. To begin with, it is instructive to plot $w(a)$ for different choices of parameters. If we first consider the models with a positive quadratic term (depicted in Fig.~(\ref{slowrolltrajectory})), we see that, for a fixed ${\tilde w}_0$, the evolution of the different solutions all converge to a single trajectory and seem to have the same slope as $a$ decreases. 
This is qualitatively different for what happens for models with a negative quadratic term; in Fig.~(\ref{hilltoptrajectory}), we plot a few models with the same ${\tilde w}_0$ but with a range of different slopes, as $a$ decreases (see also \cite{Dutta:2008qn} for further excellent discussion on the $w(a)$ evolutionary trajectories of hilltop models). This is a strong indication of what one should expect when mapping each of these theories onto the $({\tilde w}_0, {\tilde w}_a)$ plane. The positive quadratic model will be locked in a narrow line along this plane as its parameters are varied, while the negative quadratic model will sweep out across the plane as its parameters are varied because these variations will create substantially different $w(a)$ trajectories. 

\begin{figure}
    \centering
    {\includegraphics[width=\columnwidth]{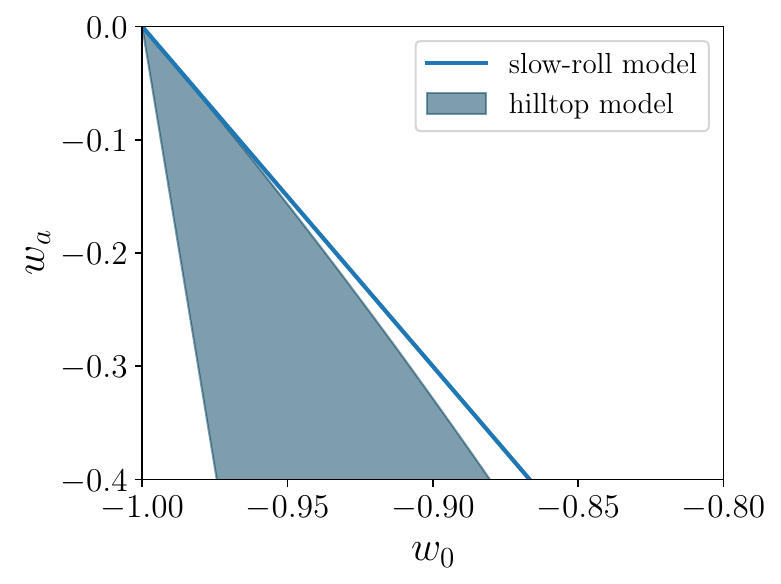}}
    \caption{${\tilde w}_0\equiv w(a=1)$ and ${\tilde w}_a\equiv-dw/da(a=1)$  assuming matter domination for both the slow-roll/positive quadratic model and the hilltop/negative quadratic model. The slow-roll model was randomly sampled for parameters $mt_0\in[.01,1]$ and $X\in[.1,2]$ and the hilltop model was randomly sampled for parameters $mt_0\in[.01,6]$ and $\epsilon\in[.0002,2]$. The hilltop model can arbitrarily sweep across huge swaths of the broader thawing (${\tilde w}_a < 0$) region when compared to more standard slow-roll models that only live along a narrow strip of this region given by ${\tilde w}_a \sim -3 ({\tilde w}_0+1)$.}
    \label{a1sweep}
\end{figure}

In order explore the $({\tilde w}_0, {\tilde w}_a)$ plane, we first use the analytic expressions we have derived, and sample over a range different parameter values to fill in the $({\tilde w}_0, {\tilde w}_a)$ plane. As expected, the slow-roll model lies on the line with a slope $-3$, depicted in Fig.~(\ref{a1sweep}). This model (and any others that it approximates) are locked in this narrow region of the ${\tilde w}_0, {\tilde w}_a$ parameter space. 

The hilltop version of the quadratic model, on the other hand, picks up where the slow-roll model leaves off and sweeps over the steeper ranges of the $({\tilde w}_0, {\tilde w}_a)$ plane as Fig.~(\ref{a1sweep}) shows. This model (and similarly any of the other many models that it approximates) can occupy large swathes of parameter space and in principle can sweep up to the ``phantom" line (${\tilde w}_0 < -1$), at which point more exotic physics would clearly be required.

Our analytic results for a matter dominated universe are already a clear indication that we can obtain almost any value of $({\tilde w}_0, {\tilde w}_a)$ with this simple model. Given that, under general conditions for many models of thawing quintessence, one can express the Taylor expansions of their potentials as the potential described by this simple model, this is confirmation that these models are underdetermined. In other words, many models will lead to the same observational results.

Until now we have based our reasoning on analytic expressions for the Taylor expansion of $w(a)$ in the matter era; they tell us exactly what drives the changes in slope in terms of the parameters of the microphysics theory. But, of course, the universe is no longer matter dominated so it will be important to make sure that these conclusions hold up under a more realistic scenario. To check that is the case, we numerically integrate Eq.~(\ref{ScalarEOM}) for a mixed dark matter/dark energy universe. Instead of assuming a pre-determined evolution for $a$, we solve the Friedman equation in the presence of the scalar field so that
\begin{equation}\label{heq}
    H^2(a)=\frac{1}{3}\left(\rho_{\mathrm{m0}} a^{-3}+\frac{1}{2}{\dot\varphi}^2+V(\varphi)\right).
\end{equation}
We choose initial conditions (at some suitably early time) such that  $\dot{\varphi}\simeq 0$,  and  a value of $\varphi_i$ that leads to $\Omega_{DE}\simeq.7$ and $H=1$ at the end of the integration (at $a=1$). 
We then generate $w(a)$ numerically for a wide variety of parameters and find (${\tilde w}_0, {\tilde w}_a$) for both models. 

\begin{figure}
    \subfloat[Slow-roll thawing quintessence trajectories.\label{ananumcomp}]{\includegraphics[width=\columnwidth]{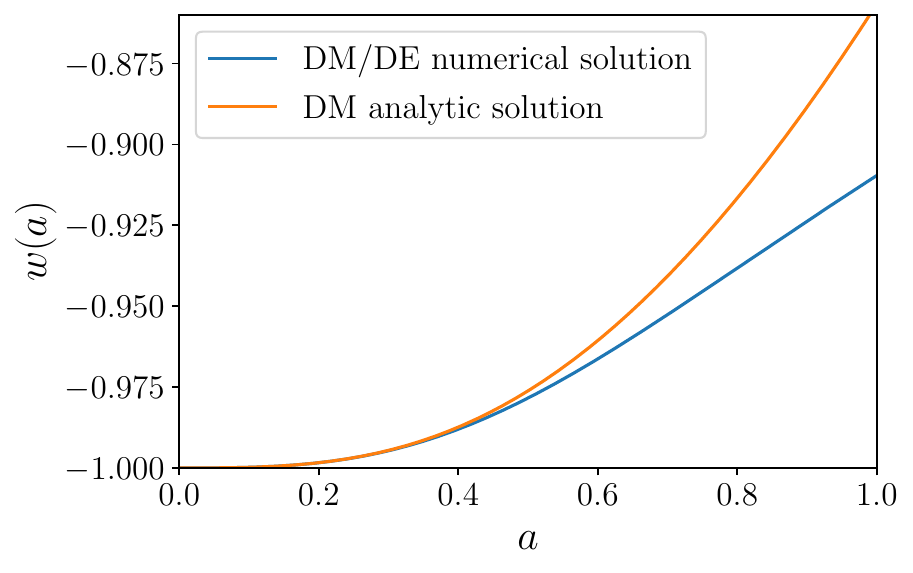}}
    \hfill
    \subfloat[Hilltop thawing quintessence trajectories.\label{ananumcomphill}]{\includegraphics[width=\columnwidth]{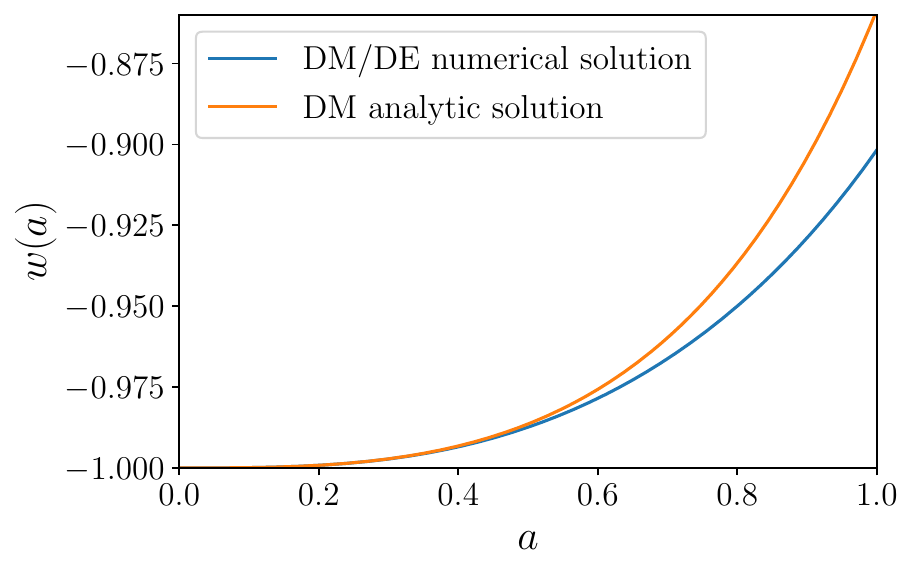}}
    \caption{$w(a)$ numerical integration for both the slow-roll/positive quadratic model and the hilltop/negative quadratic model in a mixed dark matter/dark energy universe compared to their respective matter dominated analytic solution for the same choice of parameters. In both cases, we see that the numerical solution does not evolve quite as much. This is because the Hubble friction term is enhanced when dark energy becomes a significant part of the scalar field equation of motion when compared with the matter dominated solution.}
\end{figure}

The numerical integration for the positive quadratic, slow-roll model yields a similar result: this model lives on a very thin strip of parameter space. However, the line it traces out is $ {\tilde w}_a/({\tilde w}_0+1) \sim -1.5$ (again at $a=1$), rather than $ {\tilde w}_a/({\tilde w}_0+1) \sim -3$ as in the matter dominated case. This is because in a universe with a substantial dark energy component, the Hubble friction term will be enhanced; meaning that the evolution of $w(a)$ will be suppressed when compared to the matter dominated case for the same choice of parameters (see Fig.~(\ref{ananumcomp})).
Notice also, that this numerical integration shows that for a mixed dark matter/dark energy universe, the evolution of $w(a)$ for the positive quadratic model is far more linear than the matter dominated universe. Thus, the resulting line behavior in the $({\tilde w}_0, {\tilde w}_a)$ plane will be less than the matter dominated value. This ${\tilde w}_a/({\tilde w}_0+1) \sim -1.5$ result for thawing quintessence in a mixed dark matter/dark energy universe is consistent with what was found in \cite{Garcia-Garcia:2019cvr, Scherrer:2007pu, Linder:2015zxa, Scherrer:2015tra} for similar kinds of models that can be said to represent more typical realizations of thawing quintessence.

\begin{figure}
    \centering
    {\includegraphics[width=\columnwidth]{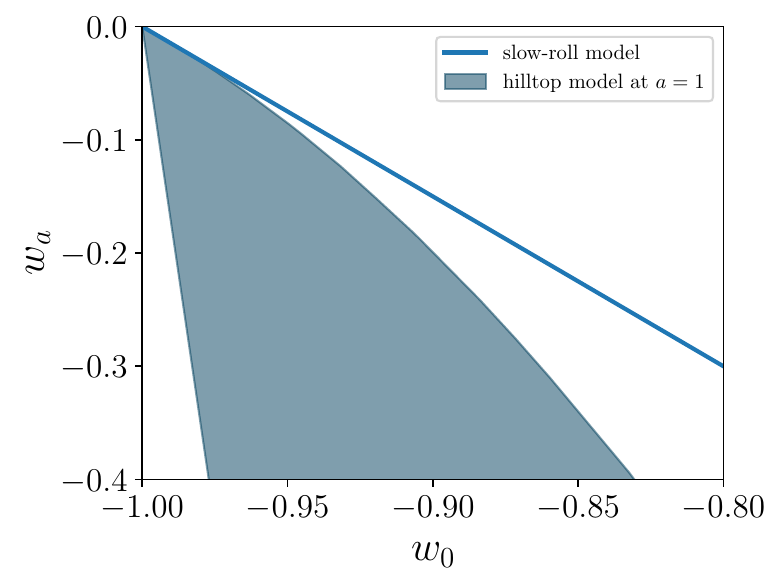}}
    \caption{This is the result for ${\tilde w}_0\equiv w(a=1)$ and ${\tilde w}_a\equiv-dw/da(a=1)$ determined after numerically integrating Eq.~(\ref{heq}) for a universe with $\Omega_{DE} \simeq .7$ for both the slow-roll/positive quadratic model and the hilltop/negative quadratic model. Both models are qualitatively similar in their behavior, with the slow-roll model now living on ${\tilde w}_a \sim -1.5 ({\tilde w}_0+1)$ while the hilltop model takes values roughly between 
    $-1.5({\tilde w}_0+1) \lesssim {\tilde w}_a \lesssim -10({\tilde w}_0+1)$ and sweeps across the parameter space.}
    \label{a1numsweep}
\end{figure}

The numerical integration for the negative quadratic, hilltop model in a mixed dark matter/dark energy universe also yields a similar result when compared with the analytic solution for a matter dominated universe: this model sweeps out a wide swath of the $({\tilde w}_0, {\tilde w}_a)$ plane. As with the positive quadratic model, the friction term is enhanced meaning that the evolution of $w(a)$ is suppressed when compared with the matter dominated case for the same choice of parameters (see Fig.~(\ref{ananumcomphill})).
Here, the evolution of $w(a)$ for the negative quadratic model is still non-linear, but the evolution is not quite as steep. Thus, we expect the resulting area that the model sweeps in the $({\tilde w}_0, {\tilde w}_a)$ to be somewhat smaller to the matter dominated case. Here, we find, for this numerical evolution and $mt_0\in[.01,6]$, $-1.5 \lesssim {\tilde w}_a/({\tilde w}_0+1) \lesssim -10$ (Fig.~(\ref{a1numsweep})).

Thus we have shown, both analytically and with numerical solutions, that we can generate an incredibly broad family of possible behaviours for the equation of state, $w(a)$. Specifically, we have shown that we can get nearly any arbitrary value of ${\tilde w}_0$ and ${\tilde w}_a$ with a quadratic model for the dark energy potential, both analytically in a matter dominated universe and numerically in a mixed dark matter/dark energy universe thought to be a good description of our current universe. This means that our conclusions implying a significant underdetermination of the microphysics underlying dark energy with respect to the observables ${\tilde w}_0$ and ${\tilde w}_a$ also hold in a realistic description of the universe, as these numerical solutions for the quadratic model will similarly map many distinct dark energy models onto the exact same regions of the $({\tilde w}_0, {\tilde w}_a)$ plane just as well as the analytic solutions in a matter dominated universe. Furthermore, it is also worth emphasizing that these results can be understood as being broadly consistent with some other studies in the literature that have considered the $({\tilde w}_0, {\tilde w}_a)$ parameterization from different perspectives than the one we have adopted here. For example, \cite{Scherrer:2015tra} takes the CPL parameterization as a starting point in order to determine which scalar field models can reasonably correspond to it and finds that a number of potentials $V(\varphi)$ are consistent with ${\tilde w}_a \sim -1.5 ({\tilde w}_0+1)$ because many potentials will exhibit sufficiently linear behavior when $\varphi$ only rolls over a small enough region of the potential. Another interesting study \cite{Huterer:2006mv} adapts the flow equation formalism from inflation to a very generic description of quintessence and also concludes that quintessence models can be found in many regions in the $({\tilde w}_0, {\tilde w}_a)$ plane outside the typical freezing and thawing bounds; where they calculate $({\tilde w}_0, {\tilde w}_a)$ from the best-measured principal components (PCs) (see \cite{Huterer:2002hy}) of the $w(a)$ trajectory.

\section{Fitting $w_0$ and $w_a$}\label{sec:surveys}

We do not directly measure ${\tilde w}_0$ and ${\tilde w}_a$. In practice, as discussed above, we {\it fit} $w_0$ and $w_a$ over a range of redshifts. The parameter space depicted in Fig.~(\ref{paramspace}) is determined in this way. Typically one has a range of distance measurements -- such as the angular diameter distance or luminosity distance of structures or objects -- which are integrals of the expansion rate and which, in turn, are a function of the energy density of dark energy. One then infers the properties of the dark energy component from how well a given model fits the distance measurements.

There are a number of Stage IV surveys that have been proposed to further our understanding of dark energy \cite{EUCLID:2011zbd,LSSTDarkEnergyScience:2012kar,SKA:2018ckk}. To simplify, we emulate a typical stage IV survey and assume that the results are measurements of the Hubble parameter as a function of redshift. Our conclusions would be no-different if we had used distance measurements themselves. As an example, we take the redshift range and forecast uncertainties in \cite{Font-Ribera:2013rwa}. 

The model we need to fit is the the Friedman equation in the form 
\begin{equation}\label{hfit}
    H^2(a)=H_0^2\left[\Omega_{\mathrm{m}} a^{-3}+\left(1-\Omega_{\mathrm{m}}\right) e^{3 w_a (a-1)} a^{-3(1+w_0+w_a)}\right].
\end{equation}
And we approximate the negative logarithm of the likelihood as 
\begin{equation}\label{xifit}
-2\ln {\cal L}\simeq\sum_i \frac{(H_{obs}(z_i)-H(z_i))^2}{\sigma_{i}^2},
\end{equation}
where $H_{obs}(z)$ is the observed $H$, $H(z)$ is the computed $H$ from Eq.~(\ref{hfit}) as a function of the $w_0$ and $w_a$ parameters, and $\sigma_{i}$ refers to the uncertainty at the redshift bin $i$.

Assume now that the observed $H(a)$ corresponds to a particular model of thawing quintessence for which we have full numerical solutions. Let us then generate numerical evolutions for $w(a)$ and determine the best $w_0, w_a$ fit using Eqs.~(\ref{param}, \ref{hfit}, \ref{xifit}) for redshifts $z\in[.15, 1.85]$ (corresponding to the redshift bins from \cite{Font-Ribera:2013rwa}). We begin with the positive quadratic model first and then proceed to the negative quadratic model.

For the positive quadratic model, this fitting procedure produces results that are very similar to what we found earlier: the $w_0, w_a$ values for the quadratic model lie on a very narrow strip of the $(w_0, w_a)$ plane right around $w_a/(w_0+1) \sim -1.5$ (again consistent with other studies such as \cite{Garcia-Garcia:2019cvr, Scherrer:2007pu, Linder:2015zxa, Scherrer:2015tra}). Upon reflection, this should not be terribly surprising. After all, as we have seen in Fig.~(\ref{ananumcomp}), the numerical solution $w(a)$ for the positive quadratic model in a mixed dark matter/dark energy dominated universe like our own is highly linear. Thus, its mapping into the $(w_0, w_a)$ plane will not change substantially between only taking these values at $a=1$ or fitting them over a quite significant range of redshifts.

\begin{figure}
    \centering
    {\includegraphics[width=\columnwidth]{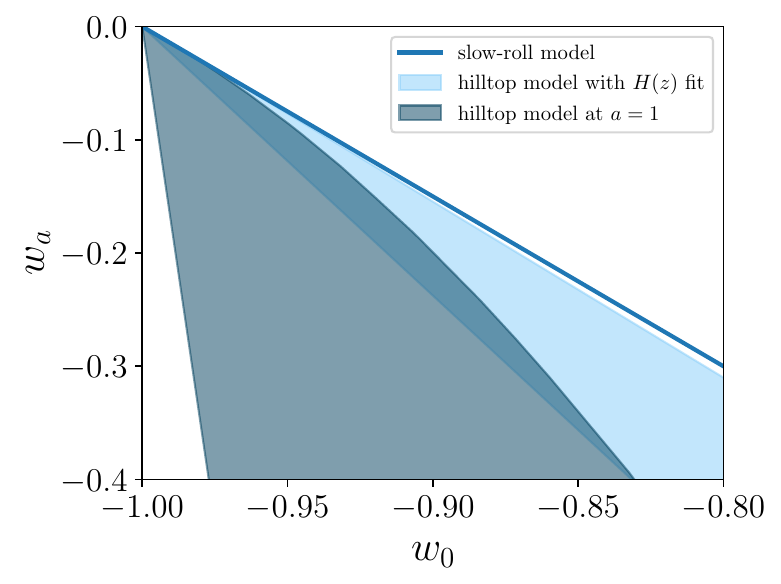}}
    \caption{The light blue shaded region depicts the result for $w_0$ and $w_a$ determined by numerically integrating $w(a)$ for the hilltop model in a universe with $\Omega_{DE} \simeq .7$ and finding the best fit for Eqs.~(\ref{hfit}-\ref{xifit}) over $z\in[.15, 1.85]$. This is overplotted against the ${\tilde w}_0\equiv w(a=1)$, ${\tilde w}_a\equiv-dw/da(a=1)$ result (dark gray shaded region) for the same hilltop model and choice of parameters depicted in Fig.~(\ref{a1numsweep}). This indicates that there is an ambiguity in how these hilltop models are represented in the $(w_0, w_a)$ plane as the exact size of the swept region (for the same choice of parameters) will sensitively depend on the range of redshifts that one fits over due to the highly non-linear evolution of $w(a)$ in these hilltop models. Fitting over a more restricted range of more recent redshifts would cause the fitted results to more closely resemble the $a=1$ results. By contrast, the slow-roll models (dark blue line) still lie in the same narrow strip as their location is insensitive to the choice of fitting procedure as their evolution is highly linear.}
    \label{hfitnumsweep}
\end{figure}

\begin{figure}
    {\includegraphics[width=\columnwidth]{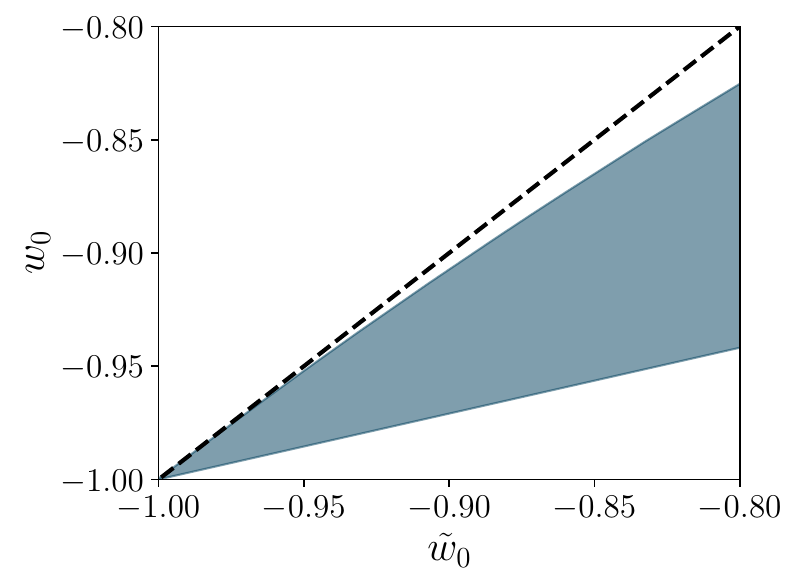}}
    \hfill
    {\includegraphics[width=\columnwidth]{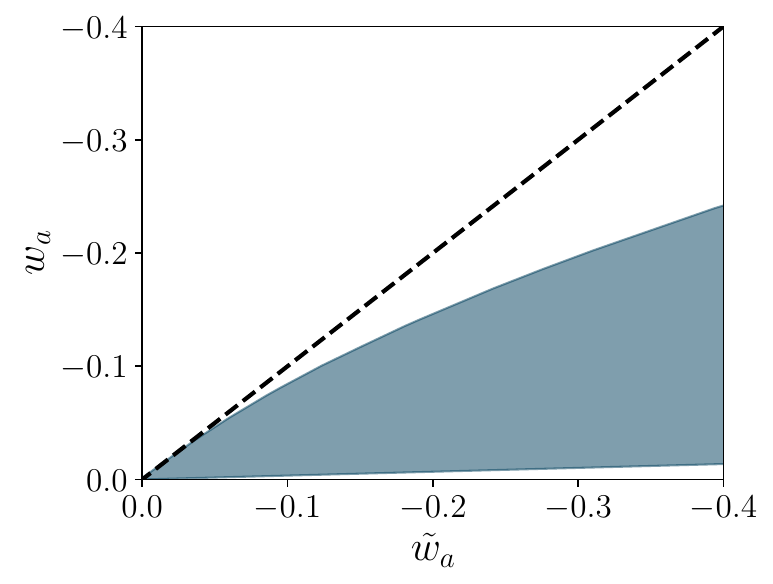}}
    \caption{Comparison between the $w_a$ and $w_0$ values depicted in the $(w_0,w_a)$ plane of Fig.~(\ref{hfitnumsweep}). This shows how the values determined by fitting $H(z)$ through Eqs.~(\ref{hfit}-\ref{xifit}) and those determined by ${\tilde w}_0$ and ${\tilde w}_a$ map onto each other. In other words, this fitting procedure squeezes the representation of the hilltop model in the $(w_0,w_a)$ plane. The exact degree to which the swept region is squeezed depends on the range of redshifts included in the fit.}
    \label{hfitcomparewaw0}
\end{figure}

The negative quadratic model, on the other hand, maps quite differently into the $(w_0, w_a)$ plane when this fitting procedure is employed. Qualitatively, we still see that various parameter choices for this model will produce a `cloud' in the parameter space (rather than a narrow line as with the positive quadratic model). However, this cloud is significantly more narrow than we saw in the $a=1$ case as it very roughly lies between $ -1.5 \lesssim w_a/(w_0+1) \lesssim -2.5$. Fig.~(\ref{hfitnumsweep}) depicts the $(w_0, w_a)$ plane for the same numerical evolution as Fig.~(\ref{a1numsweep}), except that $w_0$ and $w_a$ have been determined by fitting over $z\in[.15, 1.85]$ rather than by determining the values ${\tilde w}_0$ and ${\tilde w}_a$ at $a=1$; where as we saw this choice of parameters previously swept across $ -1.5 \lesssim {\tilde w}_a/({\tilde w}_0+1) \lesssim -10$. Fig.~(\ref{hfitcomparewaw0}) depicts how the ${\tilde w}_0$, ${\tilde w}_a$  values map onto the fitted parameters $w_0$ and $w_a$; and there we can clearly see how this fitting procedure `squeezes' the $(w_0, w_a)$ phase space that these hilltop models live in.

While this does not change the broader conclusion -- that there is a significant amount of parameter space for which the microphysics of dark energy is severely underdetermined because many distinct microphysical models can sweep significant parts of this parameter space -- this does reveal that exactly how a model of dark energy maps into the $(w_0, w_a)$ plane can potentially be sensitively dependent on somewhat arbitrary choices for fitting procedure. This surprising sensitivity can be traced to the fact that the $w(a)$ evolution in the hilltop model has significant non-linearities. Consequently, as a linear parameterization, $w_0$ and $w_a$ may not necessarily provide a good description of a dark energy model that does not feature a fairly linear evolutionary trajectory $w(a)$ (see also \cite{Scherrer:2015tra, Dutta:2008qn}). This immediately allows us to understand that the $w_0, w_a$ parameterization will not provide a good description over that range of redshifts for the hilltop model, because the vast majority of $w(a)$'s evolution is weighted at very recent redshifts (e.g.\ recall Figs.~(\ref{ananumcomphill}, \ref{hilltoptrajectory})). 

\begin{figure}
    \centering
    {\includegraphics[width=\columnwidth]{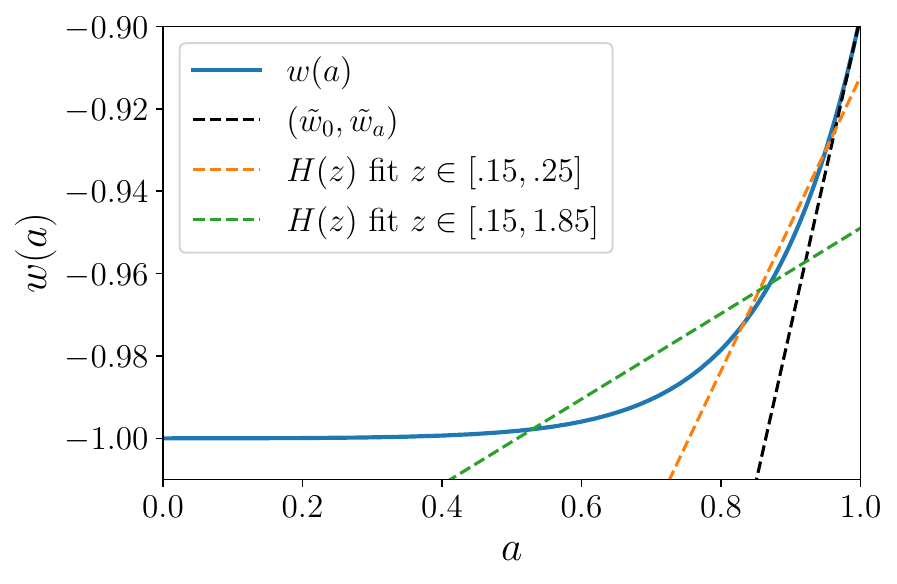}}
    \caption{Evolution of $w(a)$ compared with two different fits for the $w(a)=w_0+w_a(1-a)$ parameterization (solid blue line). The dashed green line was determined by fitting $H(z)$ through Eqs.~(\ref{hfit}-\ref{xifit}) for $z\in[.15, 1.85]$. The dashed yellow line was determined by fitting from $z\in[0.15,0.25]$. The dashed black line was determined at $a=1$. We can see that fitting over the more recent range of redshifts better captures the $w(a)$ evolution for the steeper hilltop models considered here. Consequently, we can understand why the swept area of Fig.~(\ref{hfitnumsweep}) is squeezed when fit over a large range of redshifts: the fit for $w_0$ and $w_a$ in that case does not fully capture the steeper part of the $w(a)$ trajectory that occurs at the more recent redshifts and consequently maps these models into a less steep part of the $(w_0, w_a)$ plane. The smaller and more recent the range of redshifts included in the fit, the closer that the fitted $(w_0, w_a)$ values will be to the $({\tilde w}_0, {\tilde w}_a)$ values determined at $a=1$.}
    \label{difffits}
\end{figure}

Clearly the redshift range of the data plays a significant role in the range of $w_0, w_a$ one obtains from thawing quintessence. Paradoxically, but not surprisingly,  Fig.~(\ref{difffits}) shows that, if one narrows the range of redshifts (in this case from $z\in[.15, 1.85]$ to $z\in[0.15,0.25]$)  we can see that we get a closer approximation with the ${\tilde w}_0, {\tilde w}_a$ parameterization to the full $w(a)$ evolution. Consequently, one can repeat the procedure we have done in this section, but fit over a smaller range of redshifts. One then finds that doing so will result in a wider sweeping of the $(w_0, w_a)$ plane that much more closely resembles $({\tilde w}_0, {\tilde w}_a)$ the more one restricts the range of redshifts to those where the most significant $w(a)$ evolution occurs as the fit now ``sees" the steeper part of the trajectory. The reason one might want to do this is that, as Fig.~(\ref{difffits}) clearly shows, this is a far better parameterization of the actual $w(a)$ trajectory. However, in practice, the attendant costs of doing so would require utilizing less survey data and significantly increasing the uncertainties of the observations. With all these considerations though, it is clear that there is some ambiguity in how hilltop models of the type considered here will map into the $(w_0, w_a)$ plane as different fitting procedures will result in the models sweeping different areas of the plane. Consequently, we should seek alternative parameterizations for the microphysics of dark energy that are not so dependent on such a choice of fitting procedure. 

\section{Conclusion}\label{sec:conclusion}

One of the goals of modern cosmology is to determine the microphysical model that underpins the accelerated expansion of the Universe. The most popular proposal is that it is driven by some form of dark energy which can be characterized by an equation of state. The guiding principle has been that, with current and future cosmological observations, we will be able make accurate measurements of the equation of state and, as a result, pin down the right microphysical model of dark energy. In this paper, we have focussed on a very broad class of models of dark energy -- thawing quintessence -- and showed that future observations will inevitably underdetermine the microphsyics of dark energy. 

By focusing on a widely used parametrization of the cosmological dark energy -- $w_0$ and $w_a$ -- we have shown that we can get almost any value of these parameters with a simple quadratic model for the potential. Thus, contrary to what has often been claimed in the literature, simple realizations of thawing quintessence are \textit{not} confined to a small region of the $(w_0, w_a)$ phase space. Furthermore, this does not require any highly exotic physics, unusual fine-tuning, or inordinately complicated dynamics. Essentially, one just needs a simple single field model of canonical quintessence with a quadratic potential, and in particular the hilltop flavour of this model which was first investigated in detail by \cite{Dutta:2008qn}. As we have seen, this model can find nearly any spot in the thawing region of the $(w_0, w_a)$ plane with a judicious choice of basic model parameters. We showed this using exact solutions for $\varphi$ in the matter dominated universe as well as numerically integrating the scalar field equations of motion for a mixed dark matter/dark energy universe.

This indicates that there is a significant underdetermination with respect to this model and any other model that can be placed anywhere within the $(w_0, w_a)$ plane that this quadratic model can sweep. As we demonstrated earlier, we already know of several distinct microphysical models of dark energy that can be mapped into this exact same region of the parameter space because the quadratic model approximates all of the many models that admit of a Taylor expansion of the form Eq.~(\ref{expand}) where the potential is expanded to quadratic order. Thus, these models will be indistinguishable from the quadratic model considered here from the point of view of observables in the $(w_0, w_a)$ plane.

Given that we can fill out the parameter space with the quadratic model and map between that model's predictions for the parameter space and many other models, this deflates some of the motivation for investigating models with different potentials. There could certainly be interesting or even compelling theoretical reasons or non-empirical motivations for pursuing specific models (coherence, explanatory power, aesthetics, problem-solving capabilities etc; see e.g.\ \cite{Dawid:2013maa, Wolf:2023jly, Wolf:2022yvd, Schindler2018-SCHTVI-5, Duerr:2023frq, Nyrup2015-NYRHER, Wolf:2023plp,Kuhn1977-KUHOVJ, Laudan1977-LAUPAI}). For example, there are specific potentials that are pursued for their ability to resolve outstanding fine-tuning problems \cite{Zlatev:1998tr} or that possess particularly attractive theoretical qualities such as having radiative stability due to their symmetries \cite{Frieman:1995pm}. However, it is unlikely that strictly empirical methods will single out a unique potential.

Our work does not rule out the possibility of structure formation \cite{Ferreira:2019xrr, Ruiz:2014hma, Alonso:2016suf, Wen:2023bcj, Sharma:2021ivo}, gravitational waves \cite{Baker:2017hug, Barausse:2020rsu, LISACosmologyWorkingGroup:2019mwx, Ezquiaga:2018btd, Dalang:2019rke, Wolf:2019hun, Ezquiaga:2017ekz}, fifth force tests \cite{Burrage:2017qrf, Joyce:2014kja, Will:2014kxa} etc. pointing us towards more specific microphysical realizations of dark energy. For example, non-minimal couplings could conceivably show up in any of these different types of measurements and could be used to narrow down the possible microphysical models one might consider \cite{Linder:2023klx}. Furthermore, if we are interested in learning something about the the microphysics of dark energy, we should seek alternative parameterizations of $w(a)$. As we have seen with the hilltop models considered here, the $w_0$, $w_a$ parameterization of dark energy is ambiguous in terms of how it represents certain classes of dark energy models in its parameter space due to these parameter's sensitive dependence on distance measurements. Parameterizing in terms of $(V_0, V'')$ could potentially be a more powerful approach and shed further light on the fundamental mechanism at play which is driving accelerated expansion \cite{Raveri:2021dbu, Park:2021jmi}.


\section*{Acknowledgements}

We thank David Alonso, Carlos Garcia Garcia, Martin Kunz and John Peacock  for interesting discussions. We are particularly grateful to Carlos Garcia Garcia for creating Figure 1.
PGF acknowledges support from STFC and the Beecroft Trust. WJW acknowledges support from St. Cross College, Oxford.

For the purpose of open access, the authors have applied a Creative Commons Attribution (CC BY) licence to any Author Accepted Manuscript version arising.

\bibliography{refs}

\end{document}